\documentclass[a4paper,onecolumn,11pt,accepted=2023-11-22]{quantumarticle}
\pdfoutput=1
\usepackage[utf8]{inputenc}
\usepackage[english]{babel}
\usepackage[T1]{fontenc}
\usepackage{amsmath}
\usepackage{hyperref}

\usepackage{tikz}
\usepackage{lipsum}

\begin{document}

\title{High-dimensional Encoding in the Round-Robin Differential-Phase-Shift Protocol}

\author{Mikka Stasiuk}
\affiliation{National Research Council of Canada, 100 Sussex Drive, Ottawa, Ontario K1A 0R6, Canada}
\affiliation{Institute for Quantum Computing and Department of Physics and Astronomy, University of Waterloo, N2L3G1 Waterloo, Ontario, Canada}
\author{Felix Hufnagel}
\affiliation{Nexus for Quantum Technologies, University of Ottawa, Ottawa, K1N 6N5, ON, Canada}
\affiliation{National Research Council of Canada, 100 Sussex Drive, Ottawa, Ontario K1A 0R6, Canada}
\author{Xiaoqin Gao}
\affiliation{Nexus for Quantum Technologies, University of Ottawa, Ottawa, K1N 6N5, ON, Canada}
\affiliation{National Research Council of Canada, 100 Sussex Drive, Ottawa, Ontario K1A 0R6, Canada}
\author{Aaron Z. Goldberg}
\affiliation{National Research Council of Canada, 100 Sussex Drive, Ottawa, Ontario K1A 0R6, Canada}
\affiliation{Nexus for Quantum Technologies, University of Ottawa, Ottawa, K1N 6N5, ON, Canada}
\author{Fr\'ed\'eric Bouchard}
\thanks{frederic.bouchard@nrc-cnrc.gc.ca}
\affiliation{National Research Council of Canada, 100 Sussex Drive, Ottawa, Ontario K1A 0R6, Canada}
\author{Ebrahim Karimi}
\affiliation{Nexus for Quantum Technologies, University of Ottawa, Ottawa, K1N 6N5, ON, Canada}
\affiliation{National Research Council of Canada, 100 Sussex Drive, Ottawa, Ontario K1A 0R6, Canada}
\author{Khabat Heshami}
\affiliation{National Research Council of Canada, 100 Sussex Drive, Ottawa, Ontario K1A 0R6, Canada}
\affiliation{Nexus for Quantum Technologies, University of Ottawa, Ottawa, K1N 6N5, ON, Canada}
\maketitle

\begin{abstract}
  In quantum key distribution (QKD), protocols are tailored to adopt desirable experimental attributes, including high key rates, operation in high noise levels, and practical security considerations. The round-robin differential phase shift protocol (RRDPS), falling in the family of differential phase shift protocols, was introduced to remove restrictions on the security analysis, such as the requirement to monitor signal disturbances, improving its practicality in implementations. While the RRDPS protocol requires the encoding of single photons in high-dimensional quantum states, at most, only one bit of secret key is distributed per sifted photon. However, another family of protocols, namely high-dimensional (HD) QKD, enlarges the encoding alphabet, allowing single photons to carry more than one bit of secret key each. The high-dimensional BB84 protocol exemplifies the potential benefits of such an encoding scheme, such as larger key rates and higher noise tolerance. Here, we devise an approach to extend the RRDPS QKD to an arbitrarily large encoding alphabet and explore the security consequences. We demonstrate our new framework with a proof-of-concept experiment and show that it can adapt to various experimental conditions by optimizing the protocol parameters. Our approach offers insight into bridging the gap between seemingly incompatible quantum communication schemes by leveraging the unique approaches to information encoding of both HD and DPS QKD.
\end{abstract}

\section{Introduction}

The advent of quantum key distribution demonstrated the ability to use quantum physics in public key cryptography and established one of the most studied aspects of quantum technologies. Bennet and Brassard, with the BB84 protocol, showed that encoding in two mutually unbiased bases (MUB) and randomly alternating between these encodings enables the generation of a secure random key between two parties~\cite{bennett1984quantum}. Any intervention to gain access to the random key results in detectable noise at the receiver and allows for the removal of unsecure key generation attempts~\cite{scarani2009security,bouchard2017high}. Monitoring noise has become the key element in developing a variety of quantum key distribution approaches~\cite{xu2020secure,ekert1991quantum,cerf2002security,lo2012measurement,lucamarini2018overcoming}. Recently, Sasaki \textit{et al.}~\cite{sasaki2014practical} showed a different approach to encoding random binaries in a quantum state for quantum key distribution which did not require monitoring the signal disturbance. Utilizing a quantum state in a large Hilbert space of dimension $L$, mapping random phases (therefore random phase differences) between basis vectors of the state, and a randomized interferometric measurement at the receiver lead to a fundamental bound on the mutual information between the sender and a potential eavesdropper. This introduced an entirely different approach to enforcing security in quantum key distribution~\cite{mizutani2015robustness,yin2018improved,matsuura2019refined,shan2022security}. Subsequently, several experiments used temporal~\cite{takesue2015experimental,wang2015experimental,guan2015experimental,li2016experimental,zhang2017practical,mao2017plug,wang2021round} and spatial~\cite{bouchard2018round} structures of photons to implement this protocol, also known as the Round-Robin Differential Phase Shift (RRDPS) protocol.

Advances in the preparation and measurement of high-dimensional quantum states of photons using the temporal and spatial degree of freedom motivated implementations of quantum key distribution protocols where photons carried more than one bit each. This proved beneficial at increasing secret key rates and noise tolerance~\cite{nikolopoulos2006error,mower2013high,bunandar2015practical,ding2017high,islam2017provably,bouchard2018experimental,ecker2019overcoming,vagniluca2020efficient,da2021path,bouchard2021achieving}. Notably, the issue of noise tolerance is of particular interest in the context of satellite-based QKD~\cite{sidhu2021key,gundogan2021topical,lu2022micius,sidhu2023satellite,islam2022finite}. In this work, we explore the possibility of increasing the encoding space of the secret keys in the RRDPS setting. We show that a high-dimensional quantum key distribution with and without monitoring signal disturbance is possible. Our approach both exploits the
previously untapped potential of the original RRDPS encoding scheme and offers an avenue to explore the connection between the RRDPS and BB84 protocols. We structure the paper as follows. We first introduce the high-dimensional RRDPS (HD-RRDPS) protocol. A simple sketch of the security analysis is presented. We then carry out a proof-of-principle experiment to investigate the performance and limitations of our scheme in a practical setting. Our experiment exploits the orbital angular momentum (OAM) degree of freedom of single photons, which has been demonstrated as an invaluable testbed to investigate quantum communication schemes. However, we note that our protocol can also be implemented in other degrees of freedom, such as time-bins, given appropriate measurements~\cite{brougham2013security}. Finally, in the discussion, we discuss adaptations to protocol parameters that enable better performance of the HD-RRDPS protocol for varying amounts of channel loss and characterize the circumstances under which the fundamental gap between the RRDPS and BB84 protocols can be closed. This conceptual gap can be seen both in the encoding scheme and in the security analysis for each protocol.

\begin{figure*}[t!]
	\centering
		\includegraphics[width=0.9\textwidth]{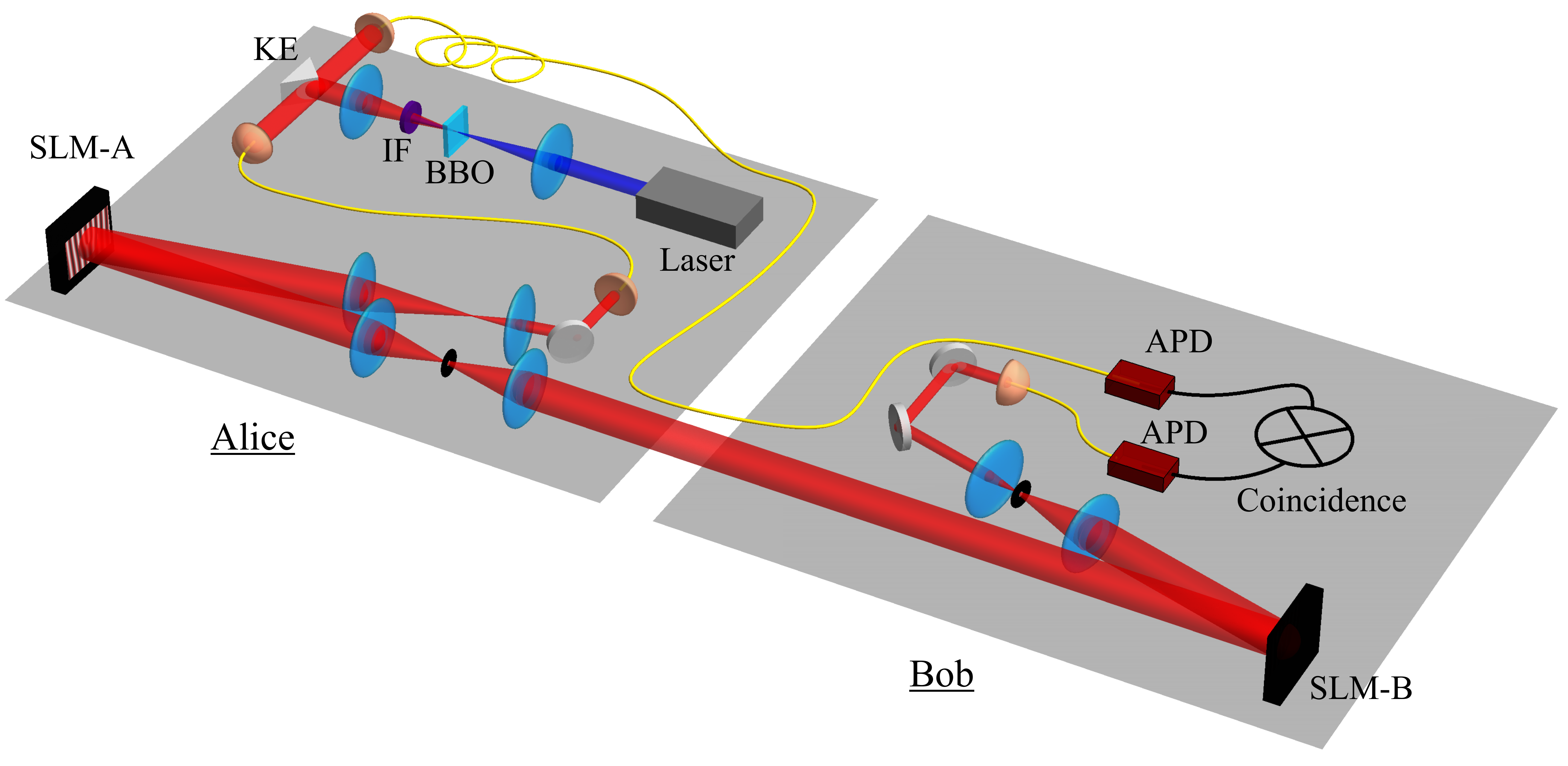}
	\caption{\textbf{Experimental Setup}. Alice generates heralded single photons at a center wavelength of 810~nm coupled to a single-mode fiber (SMF) via spontaneous parametric downconversion (SPDC) by pumping a barium borate (BBO) crystal with a diode laser at a center wavelength of 405~nm. The signal and idler photons are then spatially separated using a knife-edge (KE) mirror. We note that the pump laser is filtered using an interference filter (IF). The idler photon is subsequently used to gate Bob's measurement using coincidence measurement. Upon exiting the SMF, the signal photon is sent to a sequence of 4-f lens systems and a spatial light modulator (SLM). Alice's prepared state, $| \psi \rangle$, is encoded by imprinting the appropriate phase and intensity profile onto the signal photon using a holographic technique with SLM-A. The encoded photons are then sent to Bob's stage, where a second SLM (SLM-B) is used to measure the MUB elements, i.e. $|\varphi_m^{(d)}\rangle$. Finally, the signal and idler photons are measured using single-photon avalanche photodiode (APD) detectors and coincidence measurement.} 
	\label{fig:exp}
\end{figure*}

\section{Protocol}

We introduce the HD-RRDPS protocol as an extension of the established RRDPS protocol to a larger alphabet. In the original protocol, a superposition of pulses with randomly assigned phases of 0 or $\pi$ resulted in constructive or destructive interferences at the measurement stage. The natural extension of this encoding scheme is achieved by considering a MUB in larger dimensions. The formal key generation of the HD-RRDPS protocol is presented below. 

(1) Alice prepares a state $|\psi \rangle$ consisting of a superposition of $L$ modes which determines the dimension of the Hilbert space. In the time domain, this corresponds to a packet of $L$ pulses. Each mode is modulated by a phase $2\pi k_j/d$, where $k_j \in \{0, 1, 2, ..., (d-1)\}$, and the parameter $d$ is called the encoding dimension and satisfies $2 \leq d < L$, i.e.,
\begin{eqnarray}
|\psi\rangle = \frac{1}{\sqrt{L}} \sum_{j=1}^{L} e^{i\frac{2\pi k_j}{d}} |j\rangle.
\end{eqnarray}

(2) Upon receiving the signal state from Alice, Bob randomly selects a subset of $d$ modes out of the $L$ total modes, i.e. $\{ |j_0\rangle, |j_1\rangle, ..., |j_{(d-1)}\rangle \} \subset \{ |1\rangle, |2\rangle, ..., |L\rangle \}$, where we also have that $j_0 < j_1 < ... < j_{d-1}$. After selecting the $d$-dimensional subset, Bob performs a measurement in the MUB given by ${\{ |\varphi_m^{(d)}\rangle; m\in\{0, 1, ..., d-1\} \}}$, where 
\begin{eqnarray}
|\varphi_m^{(d)}\rangle = \frac{1}{\sqrt{d}} \sum_{n=0}^{d-1} e^{i \frac{2 \pi m n}{d}} |j_n\rangle.
\end{eqnarray}

The outcome of the MUB measurement is used to generate the raw key, $m$, which is attributed to a measurement of the state $|\varphi_m\rangle$. Moreover, Alice and Bob only keep the measurement outcomes where the state received by Bob is an element of the MUB that is being measured, i.e. ${\frac{1}{\sqrt{d}} \sum_{n=0}^{d-1} e^{i\frac{2\pi k_{j_n}}{d}} |j_n\rangle \in \{ |\varphi_m^{(d)} \}}$.

(3) Finally, Bob shares the values of ${\cal J}= \{j_0$, $j_1$, ..., $j_{(d-1)} \}$ with Alice. They can then form their final shared secure key by performing the standard classical post-processing consisting of error reconciliation and privacy amplification.\\

\section{Results}

A sketch of the security proof of the HD-RRDPS is presented here in the single-photon case. We follow the procedure presented in \cite{yin2018improved}. A detailed calculation can be found in Appendix~\ref{AppA}.

We consider the strategy adopted by the eavesdropper, Eve, where she implements a general collective attack given by ${U_\mathrm{Eve} |j\rangle |e_{00}\rangle = \sum_{\ell=1}^L c_{j\ell} |\ell \rangle |e_{j \ell}\rangle}$. The Holevo bound on Eve's reduced density matrix can be used to estimate the leaked information to Eve. For the protocol parameters $L$ and $d$, the bound on Alice and Eve's mutual information is given by
\begin{eqnarray}
I_\mathrm{AE}(x_1,x_2) \leq  \frac{ \zeta^{(d)} \left( \binom{L-1}{d-1} x_1 , \binom{L-2}{d-2} x_2, ... , \binom{L-2}{d-2} x_2 \right) }{ \binom{L-1}{d-1} \left( x_1 +x_2 \right) },
\end{eqnarray}
where we have defined $x_1$ and $x_2$ as non-negative parameters satisfying $x_1+x_2=1$, $\binom{n}{k}=n!/(k!(n-k)!)$ is the binomial coefficient, and

\begin{eqnarray}
&\zeta^{(d)}&(x_0^2, x_1^2, ..., x_{(d-1)}^2) \\ \nonumber
&=& -\sum_{i=0}^{d-1} x_i^2 \log_2 x_i^2 + \left(\sum_{i=0}^{d-1} x_i^2 \right)\log_2 \left( \sum_{i=0}^{d-1} x_i^2 \right).
\end{eqnarray}

\begin{figure}[t!]
	\centering
		\includegraphics[width=0.45\textwidth]{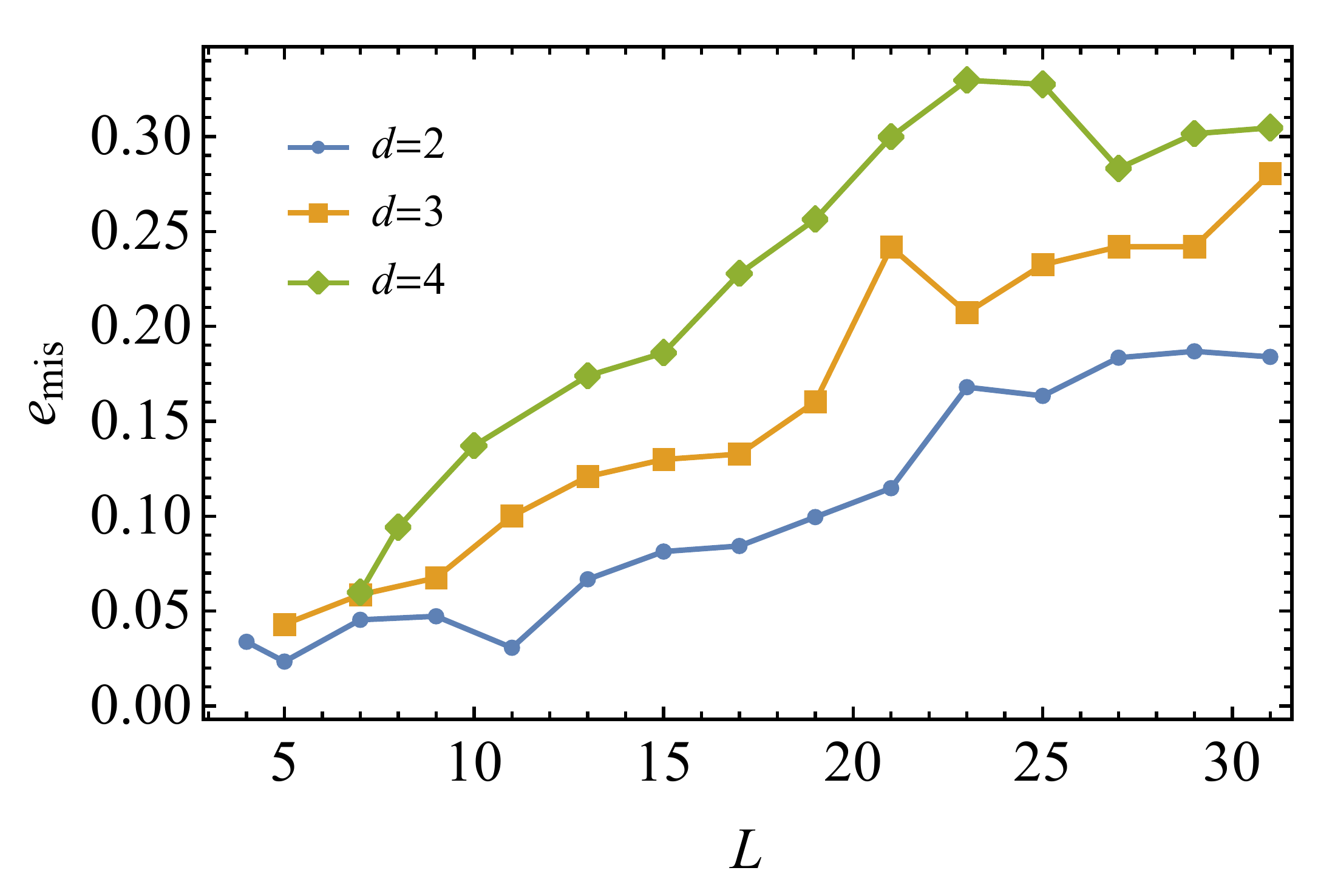}
	\caption{\textbf{Experimental characterization of mode mismatch}. Experimental averages for the mode mismatch, $e_\mathrm{mis}$ are shown for different values of the dimension, $d$, and the size of the encoding Hilbert space, $L$.} 
	\label{fig:emis}
\end{figure}

This bound arises from considering all of the paths and probabilities for Eve to obtain each of a variety of possible density matrices through a collective attack on Alice's transmitted state. We note that the expression for $I_\mathrm{AE}$ does not depend on the error rate, thus removing the requirement for monitoring signal disturbance. Nevertheless, it is possible to find a tighter bound on Alice and Eve's mutual information by determining the lower bound on the error rate of Bob's measurement in terms of $x_1$ and $x_2$. The detailed calculation is also shown in Appendix~\ref{AppA}. The secret key rate is then given by,

\begin{eqnarray}
R(E)=\log_2(d) - h^{(d)}(E) - \max_{x_1,x_2} I_\mathrm{AE}(x_1,x_2),
\end{eqnarray}
where $h^{(d)}(E){:=}-E\log_d(E/(d-1))-(1-E)\log_2(1-E)$ is the $d$-dimensional Shannon entropy. This expression of the secret key rate does not require monitoring signal disturbance. However, we can obtain the lower bound on the error rate given by
\begin{align}
    E \geq \frac{(d-1)}{d} \left( \frac{L-d}{L-1} \right) \left(\frac{x_2}{x_1+x_2} \right).
\end{align}

This inequality can be used to find a lower bound on $x_1$, i.e. $x_1^{(L)}(E)=1-E(d/(d-1))(L-1)/(L-d)$. By monitoring signal disturbance and experimentally determining the error rate $E$, an improved secret key rate is achieved, i.e.,
\begin{eqnarray}
{\cal R}(E)= \log_2(d) &-& h^{(d)}(E) \\ \nonumber
&-& I_\mathrm{AE}\left(x_1^{(L)}(E),1-x_1^{(L)}(E) \right).
\end{eqnarray}

\section{Experiment}
We perform a proof-of-principle experimental demonstration of our protocol using the OAM degree of freedom of photons, see Fig.~\ref{fig:exp}. In particular, we used the Laguerre-Gaussian (LG) modes as a basis for our OAM states. LG modes have a cylindrical symmetry and are composed of orthogonal states with a radial index, $p$, and an azimuthal index, $\ell$. In our experiment, we only use the azimuthal index, creating single photon states which carry angular momentum with a magnitude $\ell \hbar$ per photon along the propagation direction. These states have the characteristic azimuthally dependent phase $e^{i \ell \phi}$. The OAM degree of freedom is a popular approach in the application of high-dimensional QKD protocols, where their robustness has been repeatedly demonstrated in a wide range of quantum channels~\cite{sit2017high,bouchard2018quantum,sit2018quantum,hufnagel2019characterization,bouchard2019quantum,hufnagel2020investigation}.

\begin{figure}[t!]
	\centering
		\includegraphics[width=0.45\textwidth]{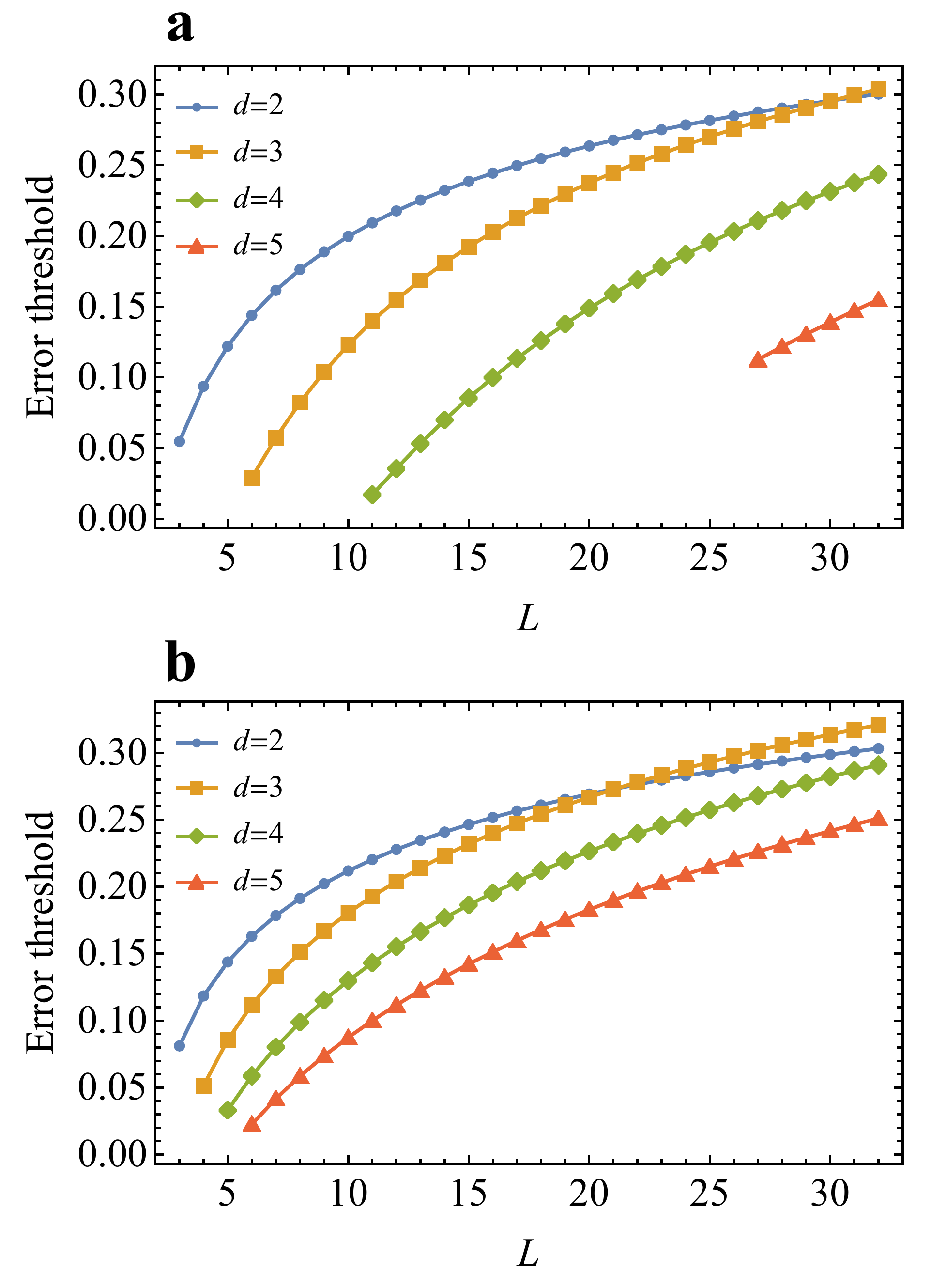}
	\caption{\textbf{Error threshold of the HD-RRDPS protocol}. Error threshold as a function of protocol parameters $L$ and $d$ (\textbf{a}) without and (\textbf{b}) with monitoring signal disturbance.} 
	\label{fig:error}
\end{figure}

Single photon pairs are produced through spontaneous parametric down-conversion (SPDC). A 360~mW Cobalt UV diode laser at a center wavelength of 405~nm is used to pump a type-I barium borate (BBO) crystal, which spontaneously produces pairs of photons, called the signal and idler, whose wavelengths are centered at 810~nm. We use a knife-edge mirror placed at the center of the beam to separate the photon pairs. The signal and idler photons are each coupled to a single-mode fiber (SMF), which selects only the Gaussian optical mode from the SPDC process. Our heralded single-photon source has a coincidence rate of 22~kHz with a 5~ns coincidence time window. The detectors each have a dark count rate of 50~Hz. The idler photon is sent directly to Bob to make a coincidence measurement jointly with the measurement of the signal photon, thus reducing the background noise in the measured data. The signal photon is used by Alice to encode the information. After exiting a fiber coupling stage, the beam is expanded using a 4-f lens system with focal lengths of 50~mm and 200~mm. The photons are sent to Alice's spatial light modulator (SLM), which imprints the desired phase to the incoming Gaussian photon to produce an OAM state. We use a phase and intensity masking technique as well as a diffraction grating to produce high-quality optical modes~\cite{bolduc2013exact}. The diffraction grating is used to send the desired phase to the first order of diffraction, which ensures that inefficient phase conversion inherent in the SLM does not result in the degradation of the mode quality. This comes at the cost of the overall efficiency of the process, as some photons will go into the other orders of diffraction. After Alice's SLM, a 4-f lens system is used to remove all other diffraction orders. We also use this 4-f system to image Alice's SLM onto Bob's SLM. The beam waist used on the SLMs was 640~$\mu$m and 650 $\mu$m for Alice and Bob, respectively. 

\begin{figure}[t!]
	\centering
		\includegraphics[width=0.45\textwidth]{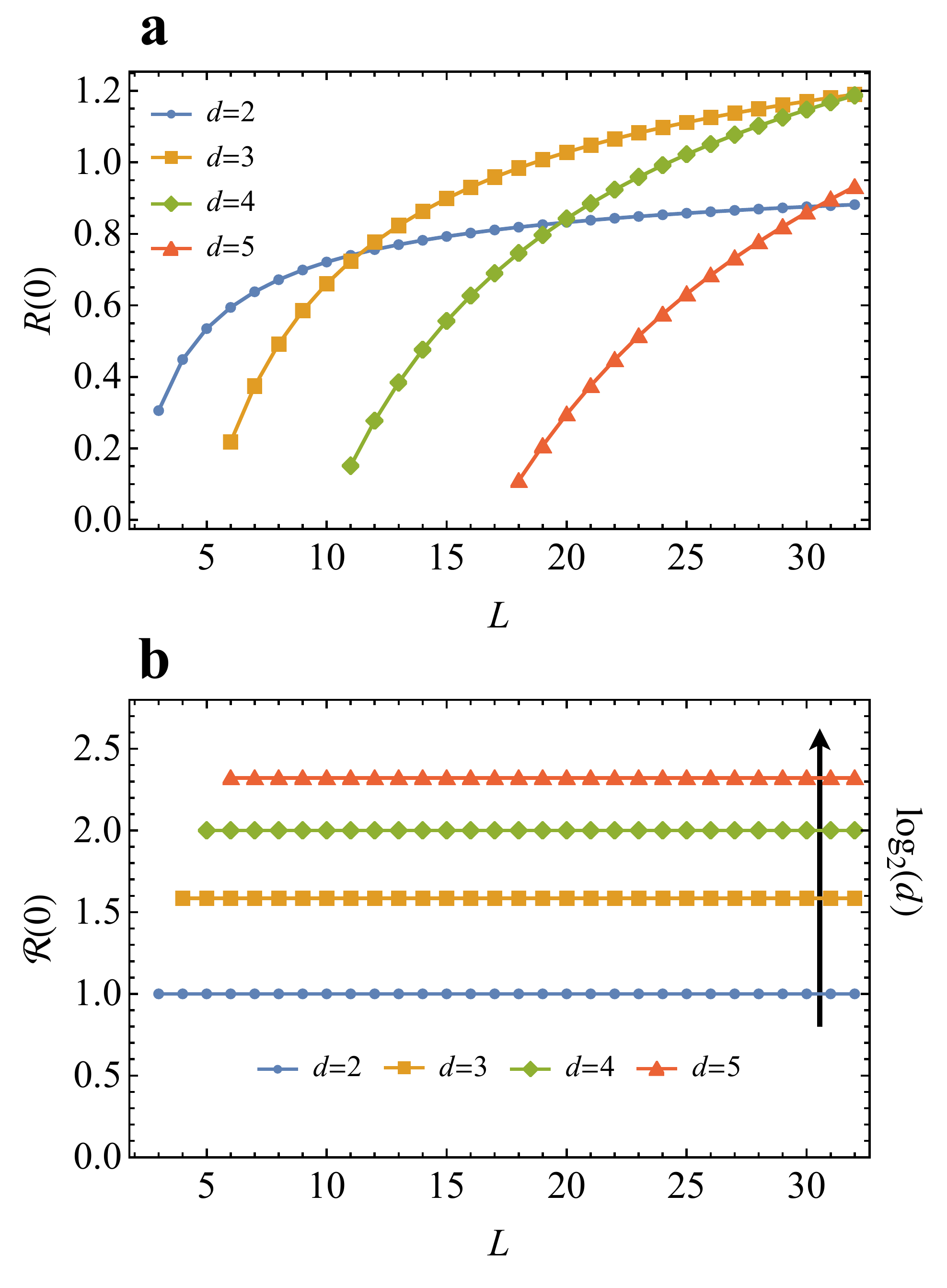}
	\caption{\textbf{Null error rate secret key rate of the HD-RRDPS protocol}. Null error rate secret key rate as a function of protocol parameters $L$ and $d$ (\textbf{a}) without and (\textbf{b}) with monitoring signal disturbance.} 
	\label{fig:errorfree}
\end{figure}

Bob measures the state of the incoming photon using the intensity flattening technique~\cite{bouchard2018measuring}. Bob displays the conjugate phase of the mode that he would like to measure on the SLM. When the incoming mode corresponds to Bob's measurement, this effectively removes the transverse phase of the incoming beam resulting in a flattened wavefront which can then be made to couple to a SMF. Before coupling the beam to the single mode fiber, Bob demagnifies the beam using a 4-f system with focal lengths 100~mm and 250~mm, respectively. Finally, the photon is sent to a single-photon avalanche photodiode (APD) detector. The coincidence measurement of the signal photon is performed with the idler photon. After the state preparation and detection, the coincidence rate is 1250~Hz for the case of Alice and Bob projecting on a Gaussian state. We note that since the idler photon is not sent to the SLM, where Alice's bits are encoded, it does not contain any information about the key sent from Alice to Bob. The resulting raw counts are converted to a mode mismatch, $e_\mathrm{mis}$, for a given dimension, $d$, and size of the encoding Hilbert space, $L$, see Fig.~\ref{fig:emis}. In the QKD protocol, the mode mismatch will result in a fixed amount of error that is independent of the channel loss. We expect that the mode mismatch will be one limiting factor in the scaling of our protocol to larger values of $d$ and $L$. Nevertheless, other degrees of freedom may result in lower mode mismatch and improved performance of the protocol~\cite{bouchard2022quantum,bouchard2023measuring}. Other types of measurements may also be considered to overcome limitations from mode mismatch~\cite{sidhu2021quantum,izumi2021adaptive,sidhu2023linear}.

\begin{figure*}[t!]
	\centering
		\includegraphics[width=0.95\textwidth]{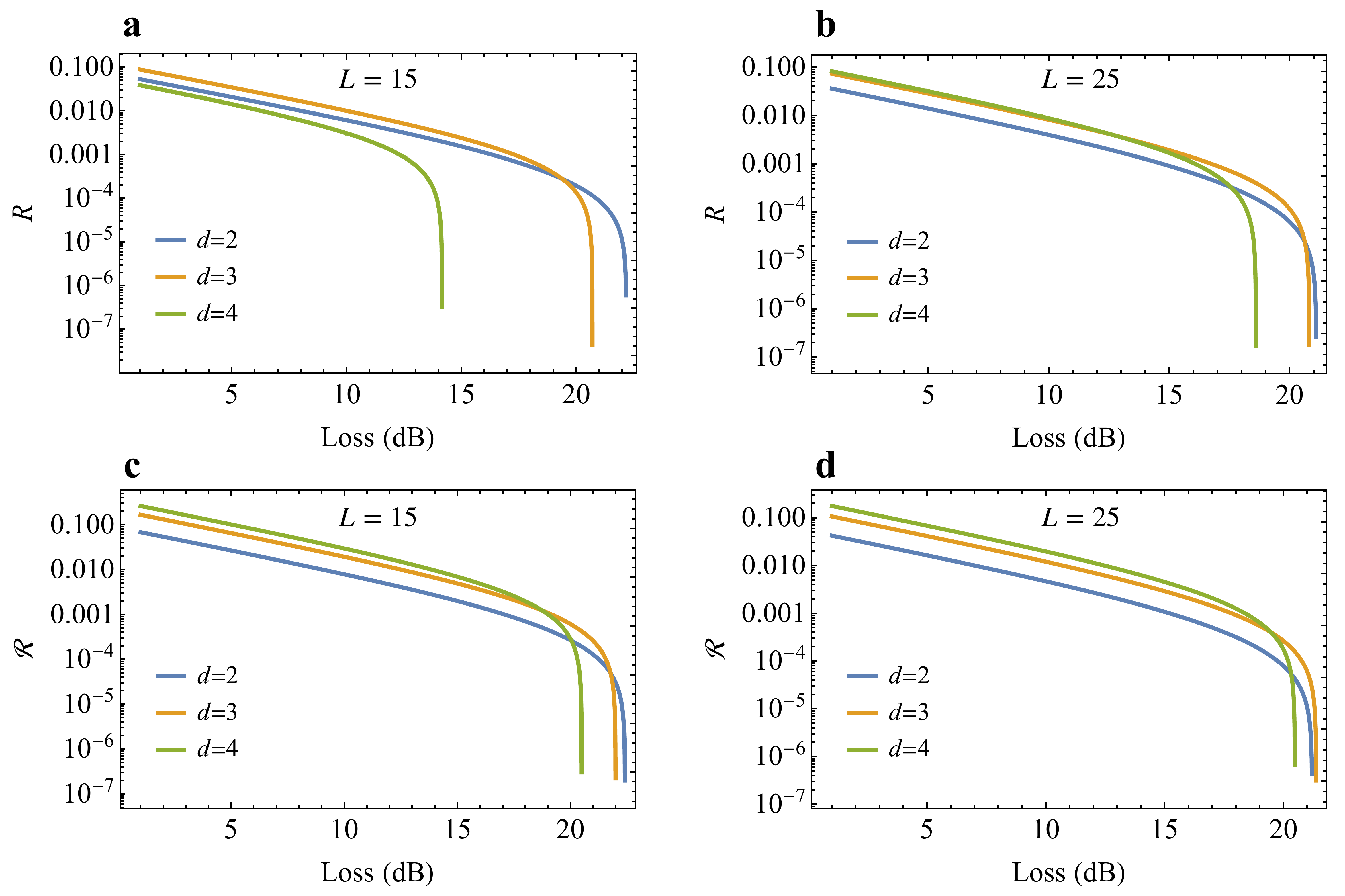}
	\caption{\textbf{Performance of the HD-RRDPS protocol}. The secret key rates as a function channel loss are shown in (\textbf{a}-\textbf{b}) without and in (\textbf{c}-\textbf{d}) with monitoring signal disturbance. The performance of the HD-RRDPS protocol is simulated using a fix value of the mode mismatch, $e_\mathrm{mis}=0.05$, for various values of the dimension, $d$, and the size of the encoding Hilbert space, $L$. In the simulation, we considered a dark count rate of $p_d=10^{-4}$.} 
	\label{fig:perfo}
\end{figure*}

\section{Discussion and Outlook}

We now discuss the performance of our HD-RRDPS protocol for various protocol parameters, i.e. $L$ and $d$, at two extreme conditions. In the first case, it is instructive to consider the case where the error rate is increased to the point where the secret key rate goes to zero. In this regime, the channel condition is noisy and we are interested in the error threshold of our protocol. In Fig.~\ref{fig:error}, we show the error threshold with and without signal disturbance for various values of $L$ and $d$. At the other extreme, we consider a condition where the level of noise is low enough to result in an null error rate secret key rate. In this scenario, we are interested in the largest achievable secret key rate. For a short quantum communication link, the key rate is limited by the single photon detectors, e.g. saturation or dead time, and when limited by the number of detected photons per second, a promising strategy involves increasing the number of secret key bits carried per photon. By doing so, it is possible to increase the overall secret key rate for the same photon detection rate. In Fig.~\ref{fig:errorfree}, we show the null error rate ($E=0$) secret key rate with and without monitoring signal disturbance for various protocol parameters.

\begin{figure*}[t!]
	\centering
		\includegraphics[width=0.95\textwidth]{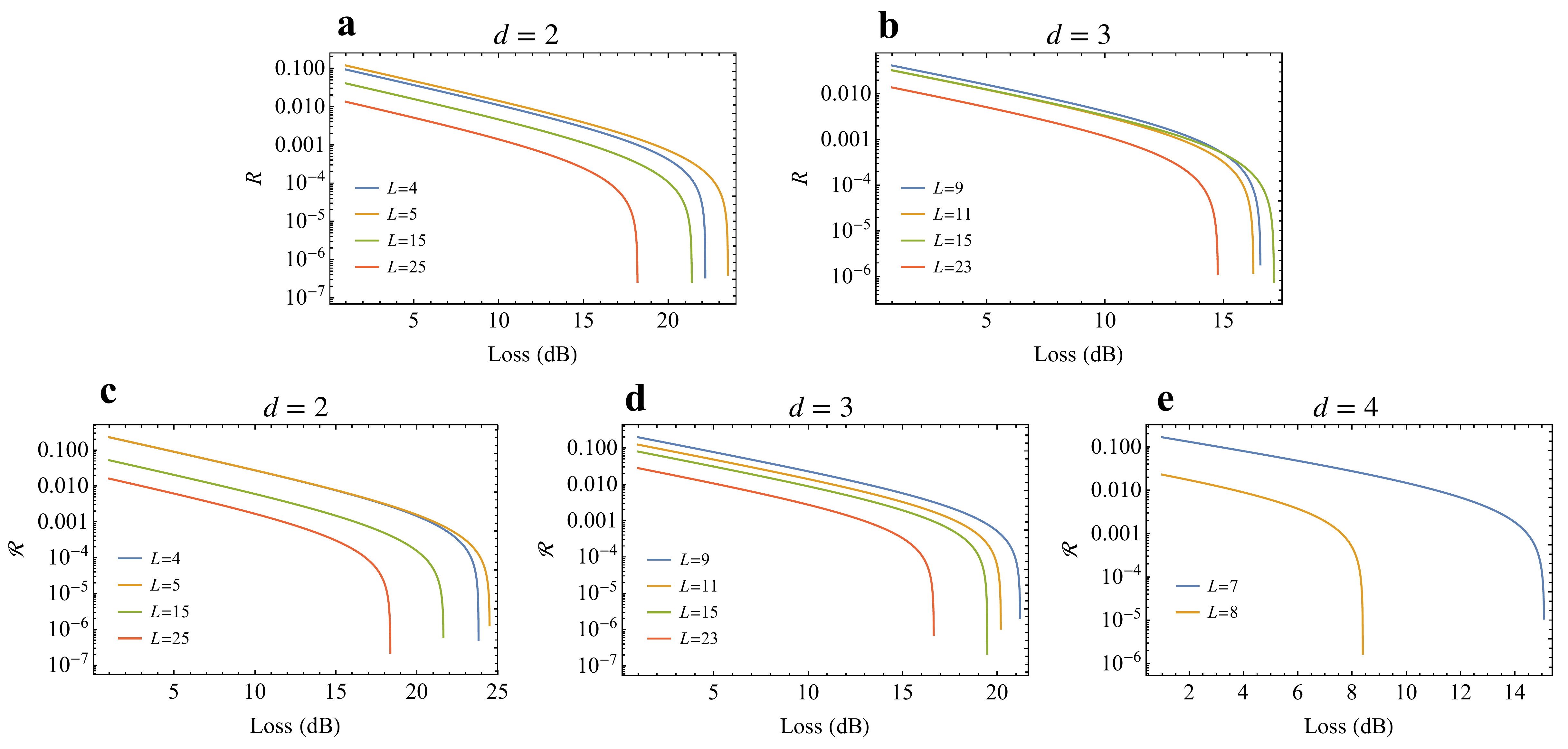}
	\caption{\textbf{Performance of the HD-RRDPS protocol from experimentally measured mode mismatch}. The secret key rates as a function channel loss are shown (\textbf{a}-\textbf{b}) without and (\textbf{c}-\textbf{e}) with monitoring signal disturbance. The performance of the HD-RRDPS protocol is simulated using the measured experimental values of the mode mismatch, $e_\mathrm{mis}$, for various values of the dimension, $d$, and the size of the encoding Hilbert space, $L$. In the simulation, we considered a dark count rate of $p_d=10^{-4}$.} 
	\label{fig:perfoexp}
\end{figure*}

In practice, QKD protocols are operated at some point between the two extreme cases considered above. Imperfections in the generation and measurement devices, and noise in the channel or the detectors will result in a non-zero error rate. We will evaluate the performance of our QKD protocol with respect to the error and secret key rates through numerical simulations. In the simulations, an overall transmission of $\eta(l)=10^{-l/10}$, where $l$ is the total loss in dB, is assigned to the communication channel. Bob detects the single-photon states using $d$ single photon detectors (SPD) with dark count rates $p_d$. For the case where Bob successfully projects the incoming state onto a $d$-dimensional MUB subspace, the single-photon yield is given by, 
\begin{eqnarray}
Y=(1-p_d)^{d-1}\left( \frac{d}{L} \eta + \left( 1-\frac{d}{L} \eta \right) d\, p_d \right),
\end{eqnarray}
where the terms $(d/L) \eta$ and $(1-(d/L)\eta)d \, p_d$ respectively correspond to the case where the signal photon is not absorbed by the channel and the case where the signal photon is absorbed by the channel and a dark count occurs. Similarly, the error rate is given by,

\begin{eqnarray}
E Y=(1-p_d)^{d-1}\left( \frac{d}{L} \eta \, e_\mathrm{mis} + \left( 1-\frac{d}{L} \eta \right) d\, p_d \right),
\end{eqnarray}
where $e_\mathrm{mis}$ is the probability that an error occurs due to a mismatch between the generation and the measurement bases. In practice, this may be the result of misalignment between the sender and the receiver, or imperfection in generation and detection devices. As a first step, we consider the performance of our protocol for a fixed value of $e_\mathrm{mis}=0.05$ that is independent of the protocol parameters $d$ and $L$. Figure~\ref{fig:perfo} demonstrates that at a low loss level, the performance of the HD-RRDPS protocol can be improved by increasing the encoding dimension, $d$, with and without monitoring signal disturbance. This advantage can persist for larger values of the size of the Hilbert space, $L$.

From our experimental measurements, we note that the mode mismatch can be dependent on the protocol parameters $d$ and $L$. This is particularly true for the case of OAM states of photons. In this case, we evaluate the performance of our protocol for varying values of the mode mismatch, $e_\mathrm{mis}$. In our proof-of-principle experiment, we characterize the value of $e_\mathrm{mis}$ for various protocol parameters such as $L$ and $d$. From these parameters, we can simulate the performance of our HD-RRDPS protocol versus channel loss, see Fig.~\ref{fig:perfoexp}.

By extending the RRDPS protocol to allow for multiple bits of raw key per photon, we gradually close the conceptual gap between differential phase-shift and high-dimensional QKD protocols. We note that our protocol can be straightforwardly extended to consider two MUB measurements rather than just one. When implementing two MUB measurements and the limiting case of $d=L$, we retrieve the high-dimensional BB84 protocol. But interestingly, in the case where $d \neq L$, we obtain a hybrid high-dimensional protocol that is a combination of the two cornerstone QKD protocols, i.e. differential phase shift and BB84. By tuning the protocol parameter $d$, one can optimize the performance of a quantum communication system under varying experimental conditions by employing the two unique information encoding schemes of HD and DPS QKD.

\begin{acknowledgments}
    Mikka Stasiuk and Felix Hufnagel contributed equally to this work. This work is supported by the High Throughput Secure Networks Challenge Program at the National Research Council of Canada and the University of Ottawa NRC Joint Centre for Extreme Photonics. We thank Duncan England, Philip Bustard, and Benjamin Sussman for insightful discussions. The authors acknowledge that the NRC headquarters is located on the traditional unceded territory of the Algonquin Anishinaabe and Mohawk people.
\end{acknowledgments}

\bibliographystyle{plain}

\onecolumn\newpage
\appendix

\section{Security proof calculation}
\label{AppA}

We present the detailed calculation of the HD-RRDPS protocol based on the procedure introduced in \cite{yin2018improved}. As mentioned in the main text, Alice prepares a state $|\psi\rangle$ given by

\begin{eqnarray}
|\psi\rangle = \frac{1}{\sqrt{L}} \sum_{j=1}^{L} e^{i\frac{2\pi k_j}{d}} |j\rangle.
\end{eqnarray}

The strategy adopted by the eavesdropper, \emph{Eve}, is a general collective attack given by the unitary transformation $U_\mathrm{Eve}$, where

\begin{eqnarray}
U_\mathrm{Eve} |j\rangle |e_{00}\rangle = \sum_{\ell=1}^L c_{j\ell} |\ell \rangle |e_{j \ell}\rangle,
\end{eqnarray}
and $|e_{j\ell}\rangle$ is Eve's ancilla state. Moreover, without loss of generality we assume that $c_{j\ell} \geq 0$ and ${\langle e_{im} | e_{jn}\rangle = \delta_{ij} \delta_{mn}}$. Upon receiving the signal state, Bob then selects a subset of modes indexed by ${ {\cal J}= \{j_0, j_1, ..., j_{(d-1)} \}}$ and performs a measurement in the MUB given by $\{ |\varphi_m^{(d)}\rangle \}$, where
\begin{eqnarray}
|\varphi_m^{(d)}\rangle = \frac{1}{\sqrt{d}} \sum_{n=0}^{d-1} e^{i \frac{2 \pi m n}{d}} |j_n\rangle.
\end{eqnarray}
We can now write the evolution of the signal state considering Eve's general collective attack and Bob's measurement, i.e.,

\begin{eqnarray}
U_\mathrm{Eve} |\psi\rangle |e_{00}\rangle \longrightarrow &\exp& \left[i \frac{2\pi k_{j_0}}{d} \right] \sum_{n=0}^{d-1} \tilde{c}_{j_0 j_n}|j_n\rangle + \exp \left[i \frac{2\pi k_{j_1}}{d} \right] \sum_{n=0}^{d-1} \tilde{c}_{j_1 j_n}|j_n\rangle + ... \nonumber \\ &+& \exp \left[i \frac{2\pi k_{j_{(d-1)}}}{d} \right] \sum_{n=0}^{d-1} \tilde{c}_{j_{(d-1)} j_n}|j_n\rangle + \sum_{\ell \notin {\cal J}} \exp \left[i \frac{2\pi k_\ell}{d} \right] \sum_{n=0}^{d-1} \tilde{c}_{\ell j_n}|j_n\rangle,
\end{eqnarray}
where we have defined $\tilde{c}_{ij}=c_{ij} |e_{ij} \rangle$. Eve's non-normalized reduced density matrix is then given by

\begin{eqnarray}
\rho_\mathrm{E} = \sum_{n=0}^{d-1} P\{\sum_{\ell=1}^L \exp \left[i \frac{2\pi k_\ell}{d} \right] \tilde{c}_{\ell j_n}  \},
\end{eqnarray}
where we have defined $P\{ |x\rangle\}=|x\rangle \langle x|$. Eve does not have any knowledge about the phases $\exp (2 \pi i k \ell /d)$ for $|\ell \rangle$ not in Bob's MUB, so averaging over all of those possible phases is equivalent to randomizing the phase of the terms $|e_{\ell j_n}\rangle$ for $\ell \not\in {\cal J}$, i.e.,

\begin{eqnarray}
\rho_\mathrm{E} = \sum_{n=0}^{d-1} P\{\sum_{p=0}^{d-1} \exp \left[i \frac{2\pi k_{j_p}}{d} \right] \tilde{c}_{j_p j_n}  \} + \sum_{n=0}^{d-1} \sum_{\ell \notin {\cal J}} c_{\ell j_n}^2 P \{ 
|e_{\ell j_n}\rangle \}.
\end{eqnarray}

In the event of Alice preparing a state $|\psi\rangle$, where we have a phase modulation with $k_{j_n} = m n$, for $n \in \{ 0,1,...,(d-1) \}$, Eve's ancilla states are given by

\begin{eqnarray}
\rho_m^{({\cal J})} = \sum_{n=0}^{d-1} P\{\sum_{p=0}^{d-1} \exp \left[i \frac{2\pi m p}{d} \right] \tilde{c}_{j_p j_n}  \} + \sum_{n=0}^{d-1} \sum_{\ell \notin {\cal J}} c_{\ell j_n}^2 P \{ 
|e_{\ell j_n}\rangle \}.
\end{eqnarray}

The mutual information between Alice and Eve is then estimated using the Holevo bound,

\begin{eqnarray}
Q^{({\cal J})} I_\mathrm{AE}^{({\cal J})} \leq &S& \left( \frac{1}{d} \sum_{m=0}^{d-1} \rho_m^{({\cal J})}  \right) - \frac{1}{d} \sum_{m=0}^{d-1} S \left( \rho_m^{({\cal J})} \right) \nonumber \\
&=& \sum_{m=0}^{d-1} \zeta^{(d)} \left( c_{j_0 j_m}^2, c_{j_1 j_m}^2, ..., c_{j_{(d-1)} j_m}^2  \right),
\end{eqnarray}
where $S$ is the von Neumann entropy, we have defined $\zeta^{(d)}(x_0^2, x_1^2, ..., x_{(d-1)}^2) = -\sum_{i=0}^{d-1} x_i^2 \log_2 x_i^2 + \left(\sum_{i=0}^{d-1} x_i^2 \right)\log_2 \left( \sum_{i=0}^{d-1} x_i^2 \right)$, and $Q^{({\cal J})}=\sum_{\ell=1}^L \sum_{n=0}^{d-1} c_{\ell j_n}^2$ is the yield of Bob projecting the signal state in the subset indexed by ${\cal J}$ and takes care of the normalization constants for Eve's reduced density matrix. Finally, Eve's information on the raw key is given by,

\begin{eqnarray}
I_\mathrm{AE} = \frac{\sum_{j_0<j_1<...<j_{(d-1)}} Q^{({\cal J})} I_\mathrm{AE}^{({\cal J})}}{\sum_{j_0<j_1<...<j_{(d-1)}} Q^{({\cal J})}} \leq \frac{\sum_{j_0<j_1<...<j_{(d-1)}} \sum_{m=0}^{d-1} \zeta^{(d)} \left( c_{j_0 j_m}^2, c_{j_1 j_m}^2, ..., c_{j_{(d-1)} j_m}^2  \right) }{ \binom{L-1}{d-1} \sum_{i=1}^L \sum_{j=1}^L c_{ij}^2 },
\end{eqnarray}
where $\binom{L-1}{d-1} = \frac{(L-1)!}{(d-1)! (L-d)!}$ is a binomial coefficient. We note that $\zeta^{(d)}(x_0^2, x_1^2, ..., x_{(d-1)}^2)$ is a concave function and we can thus use Jensen's inequality and that $\zeta^{(d)}(a \mathbf{x})=a\zeta^{(d)}(\mathbf{x})$ to simplify further Eve's information by counting all of the terms where each coefficient $c_{ij}^2$ appears,

\begin{eqnarray}
I_\mathrm{AE} \leq  \frac{ \zeta^{(d)} \left( \binom{L-1}{d-1} x_1 , \binom{L-2}{d-2} x_2, ... , \binom{L-2}{d-2} x_2 \right) }{ \binom{L-1}{d-1} \left( x_1 +x_2 \right) },
\end{eqnarray}
where we have defined $x_1 = \sum_i c_{ii}^2$ and $x_2 = \sum_{i \neq j} c_{ij}^2$. We note that $x_1$ and $x_2$ are non-negative parameters satisfying $x_1+x_2=1$ once appropriate normalization is reinstated. We can now relate the parameters $x_1$ and $x_2$ to the error rate.

We now try to further tightly bound the mutual information $I_\mathrm{AE}$ by finding a relationship between the error rate $E$ and the non-negative parameters $x_1$ and $x_2$. Bob's probability of measuring anything other than the $m$th state after Eve's measurement is,

\begin{eqnarray}
p^{({\cal J})}_m = \sum_{p \neq m} \left| \langle \varphi_p | \left( \sum_{r=0}^{d-1} \exp \left[i \frac{2\pi m r}{d} \right] \sum_{n=0}^{d-1} \tilde{c}_{j_r j_n}|j_n\rangle  + \sum_{\ell \notin {\cal J}} \exp \left[i \frac{2\pi m\ell}{d} \right] \sum_{n=0}^{d-1} \tilde{c}_{\ell j_n}|j_n\rangle \right) \right|^2,
\end{eqnarray}

\begin{eqnarray}
p^{({\cal J})}_m = \frac{1}{d} \sum_{p \neq m} \left( \left|  \sum_{n=0}^{d-1} \sum_{r=0}^{d-1} \exp \left[i \frac{2\pi(m r-pn)}{d} \right]  \tilde{c}_{j_r j_n} \right|^2 + \sum_{\ell \notin {\cal J}} \left| \sum_{n=0}^{d-1} \exp \left[i \frac{2\pi (m\ell-pn)}{d} \right] \tilde{c}_{\ell j_n} \right|^2 \right),
\end{eqnarray}

\begin{eqnarray}
p^{({\cal J})}_m = \frac{d-1}{d} \left(\sum_{n=0}^{d-1} \sum_{r=0}^{d-1} c_{j_r j_n}^2 + \sum_{\ell \notin {\cal J}}  \sum_{n=0}^{d-1}  c_{\ell j_n}^2  \right).
\end{eqnarray}

The error rate $E^{({\cal J})}$ is then given by,
\begin{eqnarray}
E^{({\cal J})}= \frac{1}{Q^{({\cal J})}} \frac{1}{d} \sum_{m=0}^{d-1} p_m^{({\cal J})},
\end{eqnarray}

\begin{eqnarray}
E^{({\cal J})}=  \frac{(d-1)}{d}  \frac{ \left(\sum_{n=0}^{d-1} \sum_{r=0}^{d-1} c_{j_r j_n}^2 + \sum_{\ell \notin {\cal J}}  \sum_{n=0}^{d-1}  c_{\ell j_n}^2  \right)}{\sum_{\ell=1}^L \sum_{n=0}^{d-1} c_{\ell j_n}^2}, 
\end{eqnarray}

The overall error is then given by,

\begin{eqnarray}
E= \frac{\sum_{j_0<j_1<...<j_{(d-1)}} Q^{({\cal J})} E^{({\cal J})}}{\sum_{j_0<j_1<...<j_{(d-1)}} Q^{({\cal J})}}, 
\end{eqnarray}

\begin{eqnarray}
E= \frac{(d-1)}{d} \frac{ \sum_{j_0<j_1<...<j_{(d-1)}}  \left(\sum_{n=0}^{d-1} \sum_{r=0}^{d-1} c_{j_r j_n}^2 + \sum_{\ell \notin {\cal J}}  \sum_{n=0}^{d-1}  c_{\ell j_n}^2  \right)}{\sum_{j_0<j_1<...<j_{(d-1)}} \sum_{\ell=1}^L \sum_{n=0}^{d-1} c_{\ell j_n}^2}, 
\end{eqnarray}

\begin{eqnarray}
E= \frac{(d-1)}{d} \left( \frac{ \binom{L-1}{d-1} \sum_{i=1}^L c_{ii}^2 + \binom{L-2}{d-2} \sum_{i \neq j}^L c_{ij}^2 + \binom{L-2}{d-1}  \sum_{i \neq j}  c_{ij}^2 }{ \binom{L-1}{d-1} \sum_{i=1}^L \sum_{j=1}^L c_{ij}^2} \right) \geq (d-1) \left( \frac{ \binom{L-2}{d-1}  \sum_{i \neq j}  c_{ij}^2 }{ \binom{L-1}{d-1} \sum_{i=1}^L \sum_{j=1}^L c_{ij}^2} \right),
\end{eqnarray}

\begin{eqnarray}
E\geq \frac{(d-1)}{d} \left( \frac{ \binom{L-2}{d-1} x_2 }{ \binom{L-1}{d-1}(x_1+x_2)} \right),
\end{eqnarray}

\begin{eqnarray}
E\geq \frac{(d-1)}{d} \left( \frac{L-d}{L-1} \right) \left( \frac{ x_2 }{ x_1+x_2} \right).
\end{eqnarray}

We note that for the case of $d=2$, we recover the results from~\cite{yin2018improved}.

\end{document}